\documentclass[a4paper,oneside,11pt]{amsart}
\usepackage{graphicx}
\usepackage[title]{appendix}
\usepackage{stmaryrd}
\usepackage{amsthm}
\usepackage{dsfont}
\usepackage{hyperref}
\usepackage{mathrsfs}
\usepackage{stmaryrd}
\usepackage[lite,initials,msc-links,nobysame]{amsrefs}
\usepackage{amsmath}
\usepackage{centernot}
\usepackage{enumerate}
\usepackage{verbatim}
\usepackage{bm}
\usepackage{amsmath,amsfonts,amscd,amssymb}
\usepackage{longtable,geometry}
\usepackage{color,soul}

\usepackage{mathtools}

\newtheorem{theorem}{Theorem}[section]
\newtheorem{lemma}[theorem]{Lemma}

\numberwithin{equation}{section}
\theoremstyle{definition}

\theoremstyle{remark}
\newtheorem{remark}{Remark}

\newcommand{\IR}{\mathbb{R}}

\newcommand{\I}{\mathbf{x}}
\newcommand{\F}{\mathbf{z}}
\newcommand{\Pl}{\mathbf{p}}
\newcommand{\G}{G}
\newcommand{\Gs}{G^*}

\newcommand{\su}{\textnormal{SU}(2)}

\newcommand{\n}{\mathbf{n}}

\newcommand{\Id}{\mathrm{Id}}
\newcommand{\qi}{\mathbf{i}}	
\newcommand{\qj}{\mathbf{j}}	
\newcommand{\qk}{\mathbf{k}}

\renewcommand{\Re}{\operatorname{Re}}

\allowdisplaybreaks

\title[Zeros of planar Ising models via flat SU(2) connections]{Zeros of planar Ising models \\ via flat SU(2) connections}

\author{Marcin Lis}{
\address{Technische Universit\"at Wien}
\email{marcin.lis@tuwien.ac.at}}

\date{\today}

\begin{document}

\maketitle

\begin{abstract}
Livine and Bonzom recently proposed a geometric formula for a certain set of complex zeros of the
partition function of the Ising model defined on planar graphs~\cite{BonLiv}. Remarkably, the zeros depend locally on the geometry of 
an immersion of the graph in the {three dimensional} Euclidean space (different immersions give rise to different zeros).
When restricted to the flat case, the weights become the critical weights on circle patterns~\cite{LisCP}. 

We rigorously prove the formula by geometrically constructing a null eigenvector of the Kac--Ward matrix whose determinant is the squared partition function.
The main ingredient of the proof is the realisation that the associated Kac--Ward transition matrix gives rise to an SU$(2)$ connection 
on the graph, creating a direct link with rotations in three dimensions. The existence of a null eigenvector turns out to be equivalent to this
connection being flat.

\end{abstract}
\section{Introduction and main result}
\begin{figure}[h]
        \includegraphics[scale=0.6]{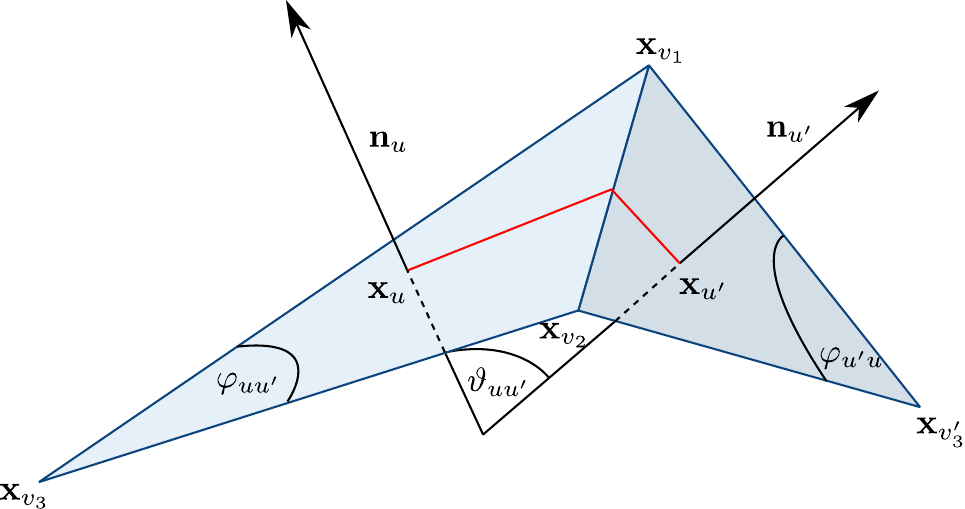}    
    \caption{Two adjacent faces of an immersion $\I$ of $G$. The triangles $\mathbf T^\I_u$ and $\mathbf T^\I_{u'}$ 
    are represented by the light blue shading. The red piecewise linear segment represents the edge $\{u,u'\}$ in~$G^*$}
    \label{fig:angles}
\end{figure}

The study of zeros of the partition function of the Ising model has a rich history going back to the seminal work of Lee and Yang~\cite{LeeYang}
who viewed it as a function of the external field and showed that all zeros are purely imaginary. This was later extended to more general spin systems~\cite{SimGri,new,LieSok}.
These results are valid irrespectively of the underlying finite graph. 

On the other hand the zeros in the coupling constans, also called \emph{Fisher zeros}, are less understood (see~\cite{FishZer} and the discussion therein).
In their recent work, Livine and Bonzom~\cite{BonLiv}, using nonrigorous methods, proposed a parametrisation of a certain class of Fisher zeros for Ising models
defined on planar graphs. Surprisingly the parametrisation concerns the geometry of embeddings (or as it turns out immersions) of the graph into the three dimensional Euclidean space. To the best of our knowledge no such 
geometric formula had been proposed before. When the embedding is locally flat, 
then locally the weights become the critical Ising weights on circle patters that were introduced by the author in~\cite{LisCP}.
In this note we give a rigorous proof of the formula of Livine and Bonzom. 

To state the result, we need to introduce the setup. To this end, let $\G=(V,E)$ be a finite connected planar graph and let $\Gs=(U,E^*)$ be its planar dual. 
We assume that $G$ is a triangulation (and hence $G^*$ is trivalent). We will identify each face $u$ of $G$ (vertex of $G^*$) with triple of vertices $\{v_1,v_2,v_3\}$ of $G$ incident to $u$.

We call an \emph{(oriented) immersion} a pair of maps $ \I: V\to \mathbb R^3$ (the immersion) and $\n: U\to \mathbb R^3$ (a normal vector) such that
for each face $u=\{v_1,v_2,v_3\}$, 
\begin{itemize}
\item the points $\I_{v_1},\I_{v_2},\I_{v_3}$ are not collinear. We denote by $\Pl_u\subset \IR^3$ their unique common plane.
\item $\n_u$ is a unit normal to $\Pl_u$ oriented in such a way that for any two neighbouring faces $u=\{v_1,v_2,v_3\} $ and $u'=\{v_1,v_2,v'_3\}$,
the bases 
\[
\I_{v_2}-\I_{v_1},\I_{v_3}-\I_{v_1},\n_u, \text{ and } \I_{v_1}-\I_{v_2},\I_{v'_3}-\I_{v_2},\n_{u'}
\] 
have the same orientation.
\end{itemize}

\begin{figure}
        \includegraphics[scale=0.25]{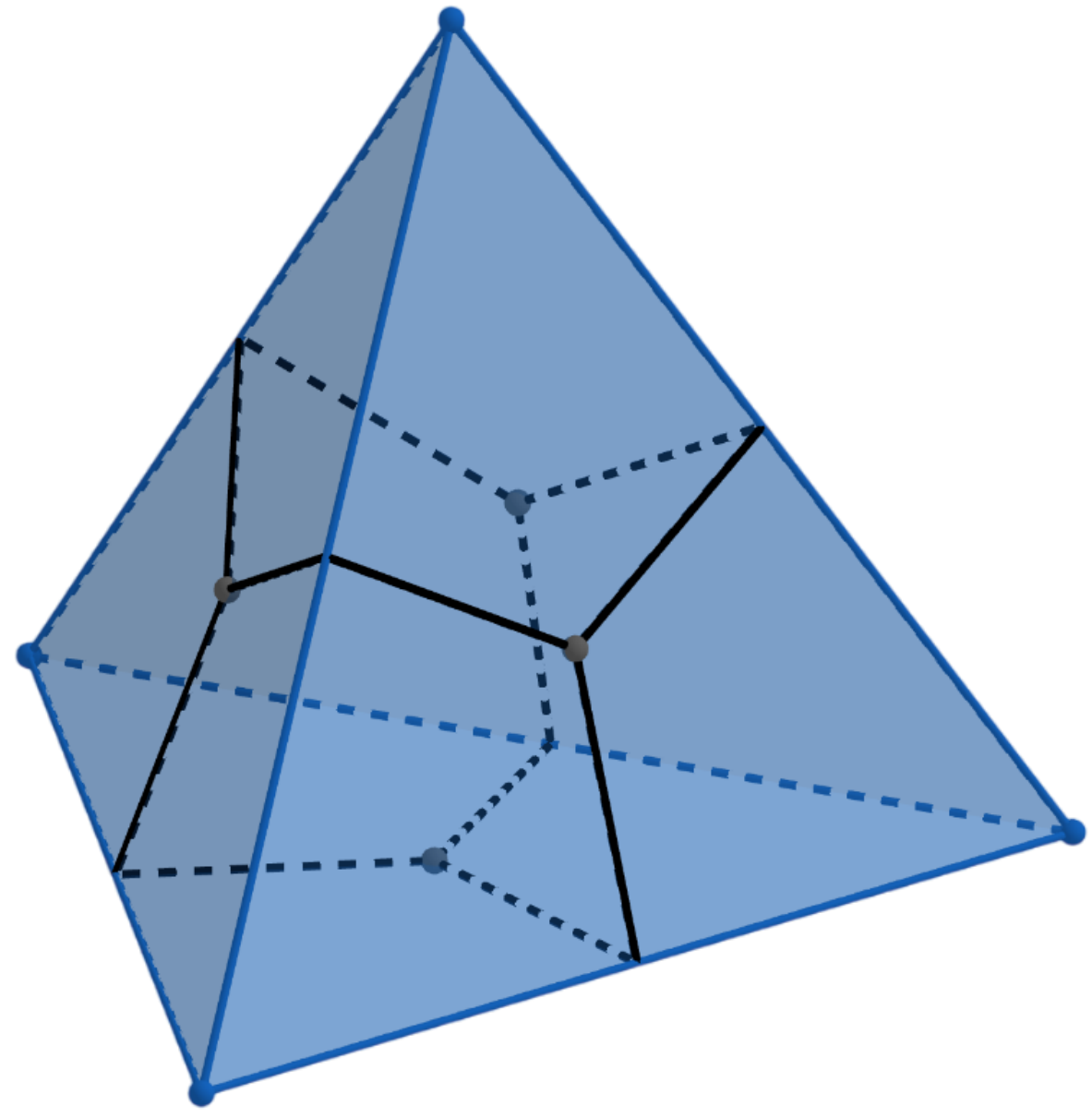}    
    \caption{An embedding of a tetrahedron $G$ drawn in blue. Its dual tetrahedron $G^*$ is drawn in black} 
    \label{fig:tetrahedron}
\end{figure}

For $\I_1,\I_2,\I_3 \in \mathbb R^d$ in general position, we denote by $\mathbf T[\I_1,\I_2,\I_3]$ the interior (in the subspace topology) of the filled triangle with vertices $\I_1,\I_2,\I_3$.
We will extend the immersion $\I$ to $U$ by assigning to each $u=\{v_1,v_2,v_3\}$ a point $\I_u$ in $\mathbf T^\I_u:=\mathbf T[\I_{v_1},\I_{v_2},\I_{v_3}]$. If $\mathbf T^\I_u$ is an acute triangle, then we make the convention of choosing $\I_u$ to be the centre of the circumcircle of $\mathbf T^\I_u$. 
We will always assume that $\I_{v_1},\I_{v_2},\I_{v_3}$ appear on the boundary of $\mathbf T^\I_u$ in the clockwise order around $\mathbf n_u$ (when looking against the direction of $\mathbf n_u$, as in Fig.~\ref{fig:angles}). The orientation condition in the definition above is equivalent to saying that that if $u'=\{v_1,v_2,v'_3\}$ is a neighbour face of $u$, 
then $\I_{v_1},\I_{v_2},\I_{v'_3}$ appear on the boundary of $\mathbf T^\I_{u'}$ in the counterclockwise order around $\n_{u'}$ (as in Fig.~\ref{fig:angles}).

Let $\mathcal G=(\mathcal V,\mathcal E)$ be a general finite graph (to distinguish from $G$ and $G^*$). We call $\omega \subset \mathcal E$ an \emph{even subgraph} of $\mathcal G$ if the degree of each $v\in \mathcal V$ in the graph 
$(\mathcal V,\omega)$ is even.
In particular $\omega=\emptyset$ is an even subgraph. We write $\Omega_{\mathcal G}$ for the set of all even subgraphs of $\mathcal G$, and given a set of weights $x: \mathcal E \to \mathbb C$, we define the generating function of even subgraphs of $\mathcal G$ by
\[
Z_{\mathcal G}(x)= \sum_{\omega \in \Omega_{\mathcal G}} \prod_{e\in \omega} x_e.
\]
By the so-called \emph{high-temperature expansion} it is classical that this is (up to an explicit multiplicative constant) the partition function of the Ising model on $\mathcal G$ with coupling constants $\mathcal J:\mathcal E\to \mathbb C$
given by $\tanh \mathcal J_e=x_{e}$. In the case when $\mathcal G$ is trivalent, the even subgraphs are simply collections of disjoint cycles, and then $Z_{\mathcal G}(x)$ is sometimes called the \emph{loop polynomial}
of $\mathcal G$ (as in~\cite{BonLiv}).

Given an \emph{embedding} $\I$ of $G$ (by which we mean an oriented immersion with the additional constraint that $\mathbf T^\I_u \cap \mathbf T^\I_{u'} =\emptyset$ for any distinct $u,u'\in U$) and two adjacent faces $u=\{v_1,v_2,v_3\}, u'=\{v_1,v_2,v'_3\}\in U$, Livine and Bonzom~\cite{BonLiv} defined the weight $y=y(\I)$ by
\begin{align} \label{def:weight}
y_{\{u,u'\}}=\exp\Big({ i \frac{\vartheta_{uu'}}{2}}\Big)\sqrt{\tan \Big(\frac{\varphi_{uu'}}2\Big)\tan \Big(\frac{\varphi_{u'u}}2\Big)},
\end{align}
where $\{u,u'\}\in E^*$, $\vartheta_{uu'}=\vartheta_{u'u}\in [-\pi,\pi]$ is the oriented angle from $\mathbf n_u$ to $\mathbf n_{u'}$ (resp.~$\mathbf n_{u'}$ to $\mathbf n_{u}$) computed clockwise when 
looking in the direction $\I_{v_1}-\I_{v_2}$ (resp.~$\I_{v_2}-\I_{v_1}$), and where $\varphi_{uu'}\in [0,\pi]$ (resp.~$\varphi_{u'u} \in [0,\pi]$)
is the unoriented \emph{(face)} angle between $\I_{v_1}-\I_{v_3}$ and $\I_{v_2}-\I_{v_3}$ (resp.~between $\I_{v_1}-\I_{v'_3}$ and $\I_{v_2}-\I_{v'_3}$), see Fig.~\ref{fig:angles}.
We note that if $u$ and $u'$ are coplanar, i.e. $\mathbf p_u=\mathbf p_{u'}$, then $\vartheta_{uu'}=0$ and the resulting positive weights were proved to be critical (on infinite tilings of the flat plane)~\cite{LisCP}.

Our main result is a rigorous proof of the main claim from~\cite{BonLiv} (see also~\cite{GarEte} for numerical evidence) generalised to the setting of immersions (the definition of the weights $y$ generalises readily) and
with a corrected sign of the $\vartheta$ angles (see~\cite{GarEte} for a discussion).  
In the case of the tetrahedron (see Fig.~\ref{fig:tetrahedron}) a proof was given already in~\cite{tetrahedron}, and the case of a double pyramid was analytically checked in~\cite{GarEte}.
\begin{theorem}[Livine--Bonzom zeros] \label{thm:main}
In the setup as above we have
\[
Z_{G^*}(y)=0.
\]
\end{theorem}

Our approach (unlike the derivation of~\cite{BonLiv} that uses a relation to spin networks~\cite{BCL}) makes use of the Kac--Ward matrix~\cite{KacWard} (or more precisely its conjugated version studied in~\cite{Lis2014,Lis2014a,LisSP,LisCP}).
Its determinant is equal to the partition function squared, and hence to prove Theorem~\ref{thm:main} it is enough to construct a vector in its kernel.
We do it by a suitable change of coordinates of the eigenvector which reveals a structure of an SU(2) connection on $G^*$. The existence of the null eigenvector turns out to be equivalent to 
this connection being flat,~i.e.~having trivial holonomy around every loop. 
Due to the relation between SU(2) and rotations of $\mathbb R^3$, this has a geometric interpretation in terms of the existence of an immersion of the graph.
As a byproduct we establish that the space of eigenvectors has two complex dimensions.
We give the details in the remainder of this note.

\begin{remark}
As described in~\cite{BonLiv} the result extends readily (with properly generalised weights) to immersions of graphs~$G$ where the dual $G^*$ has higher degrees of vertices than three. 
In that case one requires that the immersion maps the vertices lying around each face of $G$ not only to a single plane but moreover to a single circle in $\mathbb R^3$ (which for three vertices is the same condition). 
If the immersion is locally flat (meaning that the angles $\vartheta$ are locally zero) this corresponds to circle patterns with the critical weights from~\cite{LisCP}. 
\end{remark}

\begin{remark}
As mentioned in~\cite{BonLiv} the dimension of the set of zeros given by formula \eqref{def:weight} is half of the full dimension of the set of all zeros of $Z_{G^*}(x)$. It is natural to expect that a larger class of zeros
will be given by an appropriate three dimensional generalisation of the critical weights of the Ising model on \emph{s-embeddings} defined by Chelkak~\cite{chelkakICM,chelkak2020ising} which in the flat case generalise the critical weights on circle patterns.
\end{remark}

\begin{remark}
If $G^*$ has the topology of a torus or a surface of higher genus, then the determinant of an associated Kac--Ward matrix is known to be equal to the square of the partition function of even subgraphs
counted
with additional signs depending on the homology class of the even subgraph~\cite{Cimasoni}. Our proof shows that with the weights defined in the same way as above, this partition function also vanishes. 
\end{remark}

\begin{remark}
Lastly, we want to mention that as the planar Ising model is a special case of the dimer model~\cite{Dubedat,bdt,DCL}, this work also motivates the question of finding three dimensional geometric descriptions of zeros of the partition function of 
the dimer model itself.
\end{remark}

%
%
%

\section{Proof of Theorem~\ref{thm:main}}
\subsection{The Kac--Ward matrix} \label{s:KW}
Let $\mathcal G=(\mathcal V,\mathcal E)$ be a connected finite planar graph.
For two adjacent vertices $v$ and $v'$, we will write $\{v,v'\}$ for the corresponding undirected edge, and $vv'$ (resp.~$v'v$) for the directed edge from $v$ to $v'$ (resp.~from $v'$ to $v$). 
We will also write $\vec{\mathcal E}$ for the set of all directed edges of~$\mathcal G$.

\begin{figure}
        \includegraphics[scale=0.6]{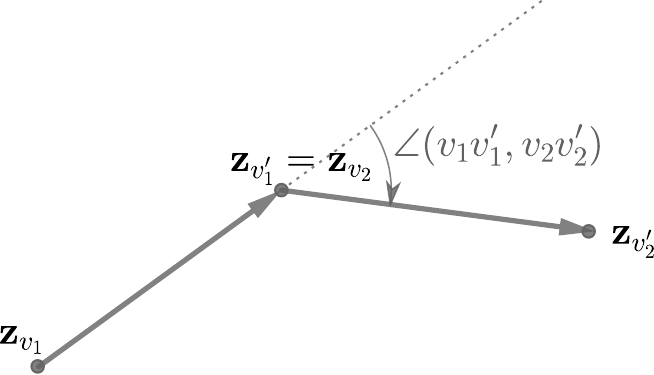}    
    \caption{The turning angle}
    \label{fig:turning}
\end{figure}

We fix a \emph{planar embedding} of $\mathcal G$, i.e.~an injective map $\F: \mathcal V\to \mathbb C \simeq \mathbb R^2$ such that 
for any $\{v_1,v'_1\},\{v_2,v'_2\}\in \mathcal E$, the open straight-line segments from $\F_{v_1}$ to~$\F_{v'_1}$ and from $\F_{v_2}$ to $\F_{v'_2}$ do not intersect.

For two directed edges $v_1v_1'$ and $v_2v_2'$, the \emph{turning angle} (under~$\F$) from $v_1v_1'$ to $v_2v_2'$ is 
\begin{align*} 
\angle(v_1v_1',v_2v_2')= \arg\Big(\frac{\F_{v_2'}-\F_{v_2}}{\F_{v_1'}-\F_{v_1}}\Big) \in (-\pi,\pi]
\end{align*}
(see Figure~\ref{fig:turning}). 
Let $\vec x: \vec{\mathcal E}\to \mathbb C$ be weights on the directed edges, and let $x: \mathcal E \to \mathbb C$ be given by
\begin{align*}
x_{\{v,v'\}}=\vec{x}_{vv'}\vec{x}_{v'v}.
\end{align*}
The \emph{(Kac--Ward) transition matrix} $\Lambda_{\mathcal G}(\vec x)$ is a matrix indexed by $\vec{\mathcal E}$ given by
\begin{align*}
[\Lambda_{\mathcal G}(\vec x)]_{v_1v'_1,v_2v'_2} = \begin{cases}
		\vec{x}_{v'_1v_1} \vec{x}_{v_2v'_2} e^{\frac{i}{2}\angle(v_1v_1',v_2v_2')}
		& \text{if } v_1'=v_2, \\
		0 & \text{otherwise}.
	\end{cases}
\end{align*}

Note the reversal of the first edge in the weight $\vec{x}_{v'_1v_1}$. 
A crucial theorem that goes back to the work of Kac and Ward~\cite{KacWard}. For streamlined proofs, see e.g.~\cite{LisSP,CCK}.
\begin{theorem}\label{thm}
\begin{align*}
{\det}(\Id -\Lambda_{\mathcal G}(\vec x)) =Z_{\mathcal G}^2(x),
\end{align*}\
where $\Id$ is the identity matrix on $\vec{\mathcal E}$.
\end{theorem}
In particular the determinant is independent of a particular choice of the embedding $\F$, and it depends on $\vec x$ only through $x$.
It immediately follows that $Z_{\mathcal G}(x)=0$ if and only if there exists $\varphi: \vec{\mathcal E} \to \mathbb C$ such that $\Lambda_{\mathcal G}(\vec x) \varphi=\varphi$.
The rest of this note is devoted to the construction of such a vector in the (slightly modified) setting of Theorem~\ref{thm:main}.

\subsection{A graph modification and planar decompositions} \label{s:mod}
The first observation is that given a finite graph $\mathcal G=(\mathcal V,\mathcal E)$ and weights $x:\mathcal E\to \mathbb C$, one can subdivide each edge $e\in \mathcal E$ into a number of subedges $e_i$ connected in a series,
and then assign weights $x_{e_i}$ in such a way that $x_e=\prod_{i}x_{e_i}$, and the resulting weighted graph will have the same generating function of even subgraphs as $\mathcal G$.

We apply this procedure to $G^*$ and divide each edge into three subedges.
More precisely, each edge $\{u,u'\}\in E^*$ is replaced by $\{u,u_1\}, \{u_1,u_2\}, \{u_2,u'\}$, where we moreover identify $u_1$ with the directed edge $uu'$, and $u_2$ with the edge $u'u$.
We denote the resulting graph by $ G^\dagger=( U^\dagger, E^\dagger)$ (where $ U^\dagger\simeq U\cup \vec E^*$) and we extend the weights $y$ defined in~\eqref{def:weight} to $E^\dagger$ by setting
\begin{align*}
y_{\{u,uu'\}}= \sqrt{\tan \Big(\frac{\varphi_{uu'}}2\Big)}, \quad y_{\{uu',u'u\}}= \exp\Big({ i \frac{\vartheta_{uu'}}{2}}\Big), \quad y_{\{u'u,u'\}}=  \sqrt{\tan \Big(\frac{\varphi_{u'u}}2\Big)}.
\end{align*}
It follows that $Z_{G^*}(y)= Z_{ G^\dagger}(y)$, which we will be crucial in the remainder of this note.

\begin{figure}
        \includegraphics[scale=0.4]{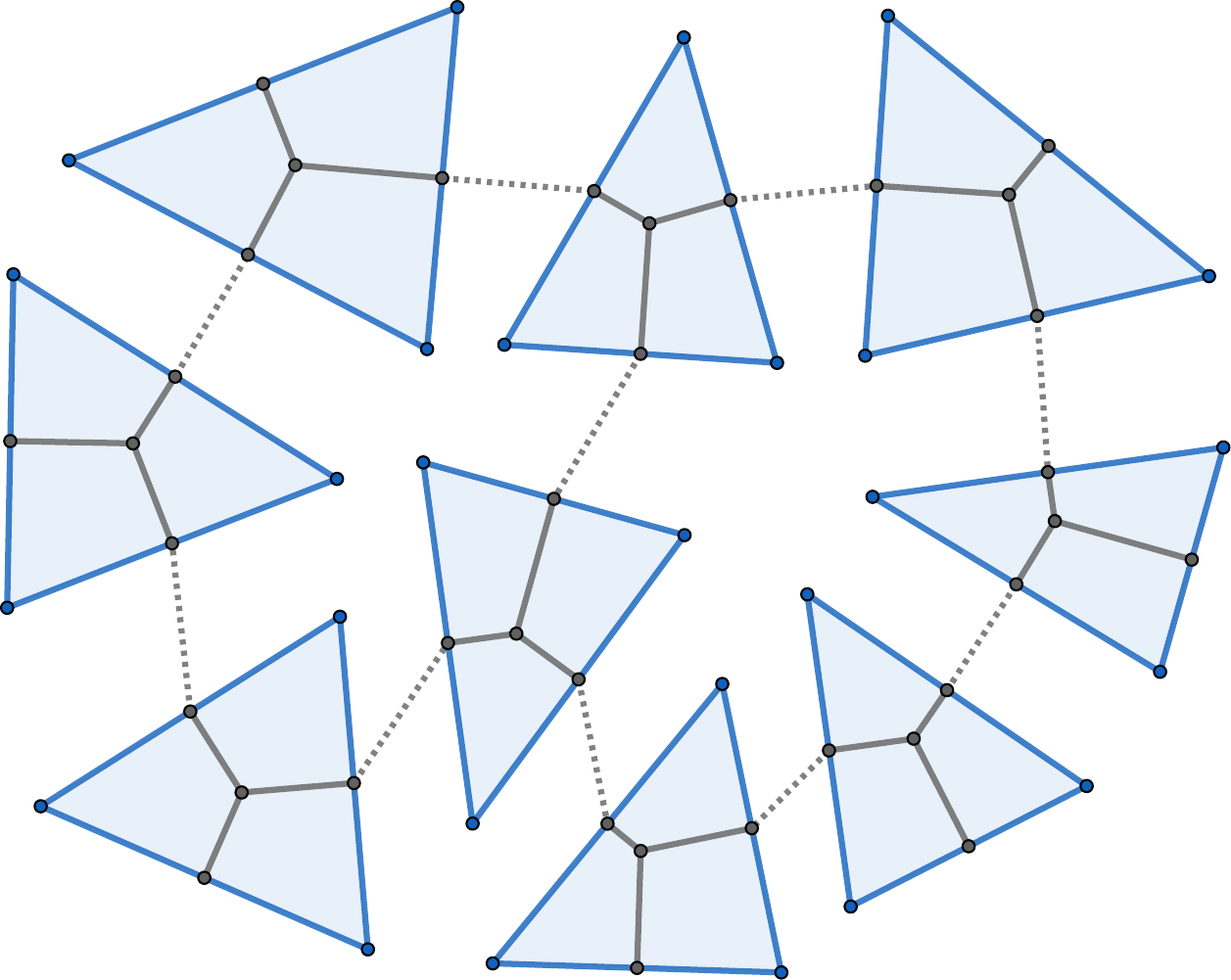}    
    \caption{A piece of a planar decomposition of an immersion}
    \label{fig:flatdec}
\end{figure}

Let $G^*$ and $\I$ be as before. A \emph{corner} of $G^*$ is a pair $uv$ of a face $u\in U$ and an incident vertex $v\in V$. We denote by $C$ the set of all corners of $G^*$.
We say that two open triangles $\mathbf T[\I_1,\I_2,\I_3]$ and $\mathbf T[\F_1,\F_2,\F_3]$ in $\mathbb R^3$ are \emph{isometric} if there exists an isometry between their closures which maps $\I_i$ to $\F_i$ for $i=1,2,3$.
In what follows we identify $\mathbb C\simeq \mathbb R^2$ with the subspace of $\mathbb R^3$ spanned by the first two vectors in the standard basis $(e_1,e_2,e_3)$ of $\mathbb R^3$.

A \emph{planar decomposition} of an immersion $\I$ consists of (see Fig.~\ref{fig:flatdec} for an illustration) 
\begin{itemize}
\item a map $\F: C\to \mathbb C$ such that for each face $u=\{v_1,v_2,v_3\}$, the triangle $\mathbf T^\F_u:=\mathbf T[\F_{uv_1},\F_{uv_2}, \F_{uv_3}] \subset \mathbb C$ is isometric to 
$\mathbf T^\I_u=\mathbf T[\I_{v_1},\I_{v_2}, \I_{v_3}] \subset\mathbb R^3$, and $\F_{uv_1},\F_{uv_2}, \F_{uv_3}$ appear in the clockwise order around the boundary of $\mathbf T^\F_u$ in the standard orientation of the complex plane, and moreover $\mathbf T^\F_u\cap \mathbf T^\F_{u'}=\emptyset$ for each pair o distinct $u, u'\in U$. We denote by $\F_u$ the image of $\I_u$ under the corresponding isometry.
\item A collection of piecewise linear and pairwise nonintersecting paths in $\mathbb C \setminus \bigcup_{u\in U} \mathbf T^\F_u$ that connect the projection of $\F_u$ on the line segment $[\F_{uv_1},\F_{uv_2}]$
with the projection of $\F_{u'}$ on the line segment $[\F_{u'v_1},\F_{u'v_2}]$ for each pair of neighbouring faces $u=\{v_1,v_2,v_3\}$, $u'=\{v_1,v_2,v'_3\}$.
\end{itemize}

We note that there always exists a flat decomposition of an immersion $\I$.
Indeed, it is enough to start with a planar embedding of $G^*$ in a big enough scale, then superimpose the isometrically imaged triangles over the vertices of $G^*$, 
and finally locally make connections via piecewise linear paths between
the midpoints of the edges of the triangles and the already existing edges of $G^*$.

From now on, for simplicity of the notation and without loss of generality, we assume that the paths defined above consist of single edges (as shown in Fig.~\ref{fig:flatdec}). 
We will extend $\F$ to $\vec E^*\cup U$ by defining $\F_{uu'}$ (resp.~$\F_{uu'}$) to be the projection of $\F_u$ on the line segment $[\F_{uv_1},\F_{uv_2}]$ (resp.~of $\F_{u'}$ on the line segment $[\F_{u'v_1},\F_{u'v_2}]$), and by taking $\F_u$ as defined above.
Note that by our convention if $\mathbf T^\F_u$ is acute, then $\F_u$ is its circumcenter, and $\F_{uu'}$ is the midpoint of $[\F_{uv_1},\F_{uv_2}]$.

This means that a planar decomposition of an immersion of $G$ gives rise to a planar embedding (as defined in Section~\ref{s:KW}) of the modified graph $ G^\dagger=( U^\dagger,  E^\dagger)$ (where $U^\dagger=U\cup \vec E^*$). We will work with the Kac--Ward transition matrix associated with this planar embedding and with the Ising weights $y$ as defined above. The corresponding Kac--Ward weights on the directed edges $\vec E^\dagger$ are chosen as follows
\begin{align} \label{eq:KWdir}
\vec y_{u(uu')}=y_{\{u,(uu')\}},\ \vec y_{(uu')u}=1, \ \vec y_{(uu')(u'u)}=\vec y_{(u'u)(uu')}= \exp\Big({ i \frac{\vartheta_{uu'}}{4}}\Big),
\end{align}
where we write $(uu')$ for a directed edge $uu'\in U^\dagger$ to avoid confusion. It is clear that $\vec y_{ww'}\vec y_{w'w}= y_{\{w,w'\}}$ for all neighbours $w,w'\in U^\dagger$, where $y$ is as above.
Hence, to prove Theorem~\ref{thm:main} it is enough to construct an invariant vector $\varphi: \vec E^\dagger\to \mathbb C$ such that $\Lambda_{G^\dagger}(\vec y)\varphi=\varphi$.

\begin{figure}
        \includegraphics[scale=0.55]{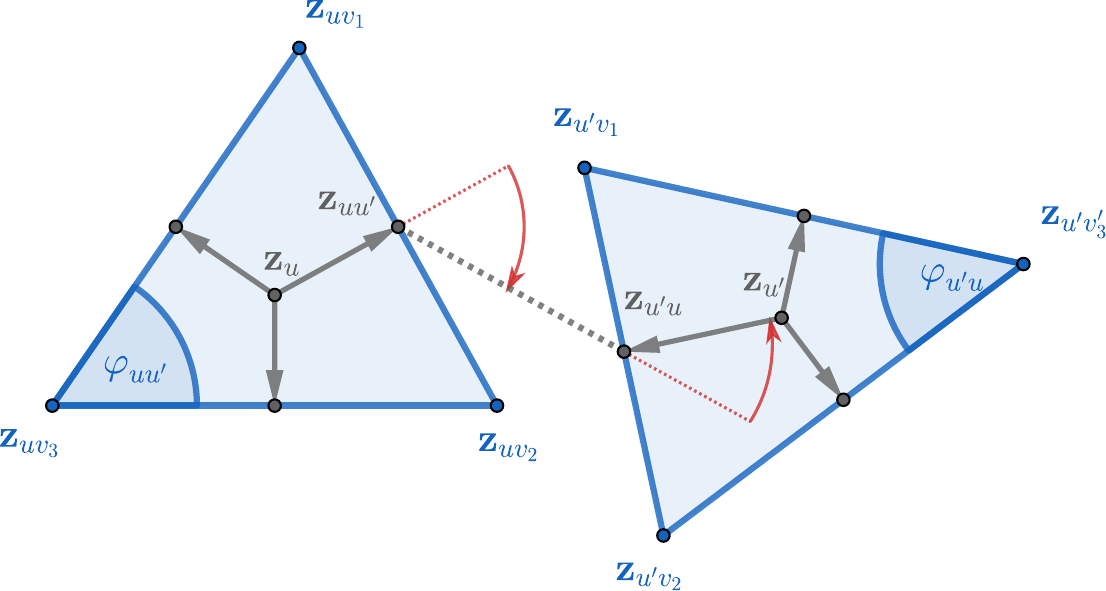}    
    \caption{Planar decomposition in the vicinity of two faces $u$ and $u'$. The angle $\alpha_{uu'}$ is equal to the sum of the two oriented red angles}
    \label{fig:transition}
\end{figure}

\subsection{A change of coordinates and an SU(2) connection}
From now on we write $\Lambda=\Lambda_{G^\dagger}(\vec y)$.
Following~\cite{Lis2014a,LisCP}, for $w\in U^\dagger$, let 
$\textnormal{Out}_w = \{ww':ww' \in \vec E^\dagger\}$, and
let $J$ be the involutive automorphism of $\mathbb{C}^{\vec E^\dagger}$ induced by the map $ww' \mapsto w'w$. As was noted in~\cite{Lis2014a}, and is easily seen from 
the definition of the Kac--Ward transition matrix, the matrix $\tilde \Lambda:=J \Lambda $ is
Hermitian, block-diagonal with blocks $\tilde \Lambda^w$, $w\in U^\dagger$, acting on the linear subspace indexed by $\textnormal{Out}_w$, and given by
\begin{align} \label{eq:JL}
\tilde \Lambda^w_{ww', ww''} = \begin{cases}
		\vec y_{ww'} \vec y_{ww''} e^{\frac{i}{2}\angle(w'w, ww'')}
		& \text{if $w' \neq w''$}; \\
		0 & \text{otherwise}.
	\end{cases}
\end{align}

Note that $\tilde \Lambda^w$ is two dimensional if $w\in \vec E^*$ and three dimensional if $w\in U$. A direct computation shows the following.
\begin{lemma} \label{lem:spectral}
When $u\in U$, the spectrum of $\tilde \Lambda^u$ is $\{-1,0,1\}$, and moreover
\begin{align*}
\tilde \Lambda^u\tilde \varphi^{u,+}= \tilde \varphi^{u,+} \quad \textnormal{ and } \quad \tilde \Lambda^u\tilde \varphi^{u,-}= -\tilde\varphi^{u,-},
\end{align*}
with $\varphi^{u,\pm}\in \mathbb C^{\textnormal{Out}_u}$ given by
\begin{align*}
 \varphi^{u,+}_{u(uu')} =\rho_{uu'} \quad \textnormal{ and } \quad  \varphi^{u,-}_{u(uu')} =\rho_{uu'}e^{-i\beta_{uu'}},
\end{align*}
where 
\begin{align*}
\rho^2_{uu'}= {|\F_{uv_1}-\F_{uv_2}|} \quad \textnormal{ and } \quad e^{i\beta_{uu'}}=\frac{\F_{uv_1}-\F_{uv_2}}{{|\F_{uv_1}-\F_{uv_2}|}}.
\end{align*}
Here $u(uu')\in \textnormal{Out}_u$, and $u=\{v_1,v_2,v_3\}$ are listed clockwise and $u'=\{v_1,v_2,v'_3\}$ are listed counterclockwise (see Fig.~\ref{fig:transition}). 
\end{lemma}
For future reference, we note that $\rho^2_{uu'}=\rho^2_{u'u}=|\I_{v_1}-\I_{v_2}|$, and elementary geometry implies that 
\begin{align} \label{eq:angeq}
 e^{i(\beta_{uu'}-\beta_{u'u})}=\frac{\F_{uv_1}-\F_{uv_2}}{\F_{u'v_2}-\F_{u'v_1}}=-e^{-i\alpha_{uu'}},
\end{align}
where
\begin{align*}
\alpha_{uu'} = \angle(uu_1,u_1u_2)+\angle(u_1u_2,u_2u'),
\end{align*}
with $u_1=uu',u_2=u'u\in U^\dagger$ (see Fig.~\ref{fig:transition}). 

Now assume that $\Lambda\varphi=\varphi$, and hence $\tilde \Lambda\varphi=J\Lambda\varphi=J\varphi$.
Since $J$ is an isometry for the Euclidean norm $\|\cdot\|$ on $\mathbb{C}^{\vec E^\dagger}$, and since $\tilde \Lambda$ is block diagonal, we can write

\begin{align*}
 \sum_{w\in U^\dagger} \|\varphi^w\|^2=\|\varphi\|^2= \|\tilde \Lambda\varphi\|^2=\sum_{w\in U^\dagger} \|\tilde \Lambda^w\varphi^w\|^2, 
 \end{align*}
where $\varphi^w$ is the restriction of $\varphi$ to the subspace indexed by $\text{Out}_w$, and $\|\tilde \Lambda^w\|$ is the Euclidean operator norm of $\tilde \Lambda^w$.
Since the blocks $\tilde\Lambda^w$ are Hermitian and their operator norm is equal to the spectral radius. If $w\in U$, then by Lemma~\ref{lem:spectral}, the spectral radius of $\tilde \Lambda^w$ is one.
On the other hand if $w\in \vec E^*$, then by~\eqref{eq:JL} and~\eqref{eq:KWdir}, $\tilde \Lambda^w$ is a $2\times2$ antidiagonal matrix with entries of modulus one, and hence also of spectral radius one.
We conclude that if $\Lambda\varphi=\varphi$, then for all $u\in U$, $\|\varphi^u\|=\|\tilde \Lambda \varphi^u\|$ and hence by Lemma~\ref{lem:spectral},
\begin{align*}
\varphi_{uw}=\varphi^u_{uw} = \xi_u^{+} \varphi^{u,+}_{uw}+ \xi_u^{-} \varphi^{u,-}_{uw}
\end{align*}
for some $\xi_u^{\pm}\in \mathbb C$. 
We therefore also have 
\begin{align*}
\varphi_{wu}=\Lambda \varphi_{wu} =J\tilde \Lambda^u \varphi^u_{wu} =J(\xi_u^{+} \varphi^{u,+}_{wu}- \xi_u^{-} \varphi^{u,-}_{wu})=\xi_u^{+} \varphi^{u,+}_{uw}- \xi_u^{-} \varphi^{u,-}_{uw}.
\end{align*}
This in particular means that the new coordinates $\xi_u=(\xi_u^{+},\xi_u^{-})$, $u\in U$, fully determine the eigenvector $\varphi$ around each $u\in U$, which in turn
determines the full eigenvector. 

We now want to see how $\xi^u$ and $\xi^{u'}$ are related
for neighbouring~${u,u'\in U}$. 
To this end, we note that when propagating the value $\varphi_{(u'u)u'}$ to $\varphi_{u(uu')}$ (resp.~$\varphi_{(uu')u}$ to $\varphi_{u'(u'u)}$) by applying $\Lambda$ (left application of $\Lambda$ moves us backwards along the directed path from $u$ to $u'$), one picks up the phase $e^{\tfrac i2(\vartheta_{uu'}+\alpha_{uu'})}$ (resp.~$e^{\tfrac i2(\vartheta_{uu'}-\alpha_{uu'})}$), and hence
\begin{align*}
\varphi_{u(uu')} = \varphi_{(u'u)u'}e^{\tfrac i2(\vartheta_{uu'}+\alpha_{uu'})} \quad \textnormal{ and } \quad \varphi_{u'(u'u)} = \varphi_{(uu')u}e^{\tfrac i2(\vartheta_{uu'}-\alpha_{uu'})}.
\end{align*} 
This together with the two displayed equations above and the explicit form of the eigenvectors from Lemma~\ref{lem:spectral} give (after factoring out $\rho_{uu'}$ from both equations)
\begin{align*}
\xi_{u'}^+-\xi_{u'}^-e^{-i\beta_{u'u}}&=(\xi_{u}^++\xi_u^-e^{-i\beta_{uu'}})e^{\tfrac i2(-\vartheta_{uu'}-\alpha_{uu'})},\\
\xi_{u'}^{+}+\xi_{u'}^{-}e^{-i\beta_{u'u}}&=(\xi_u^{+}-\xi_u^{-}e^{-i\beta_{uu'}})e^{\tfrac i2(\vartheta_{uu'}-\alpha_{uu'})}, \\
\end{align*}
which in matrix form reads 
\[
\xi_{u'}= \xi_{u} \Upsilon_{uu'},
\]
where 
\begin{align*}
\Upsilon_{uu'}&=e^{-\frac i2 \alpha_{uu'}}\begin{pmatrix}
\cos(\frac{\vartheta_{uu'}}2) & ie^{ i\beta_{uu'}}\sin(\frac{\vartheta_{uu'}}2)  \\
-ie^{- i\beta_{u'u}}\sin(\tfrac{\vartheta_{uu'}}2)  & -e^{i(\beta_{u'u}-\beta_{uu'}) } \cos(\tfrac{\vartheta_{uu'}}2)
\end{pmatrix} \\
&=\begin{pmatrix}e^{-\frac i2 \alpha_{uu'}}
\cos(\frac{\vartheta_{uu'}}2) & -ie^{\frac i2 \alpha_{uu'}+i\beta_{uu'}}\sin(\frac{\vartheta_{uu'}}2)  \\
-ie^{-\frac i2 \alpha_{uu'}- i\beta_{uu'}}\sin(\tfrac{\vartheta_{uu'}}2)  & e^{\frac i2 \alpha_{uu'}} \cos(\tfrac{\vartheta_{uu'}}2)
\end{pmatrix}.
\end{align*}
In the second identity we used \eqref{eq:angeq}. On can see that $\Upsilon_{uu'}\in \su$ as it is of the form  
\begin{align} \label{eq:su2}
\Upsilon =\begin{pmatrix} 
a +b i & c+di \\
-c+di &  a-bi
\end{pmatrix}
\end{align}
with $a,b,c,d\in \mathbb R$ and $a^2+b^2+c^2+d^2=1$. By construction, $\Upsilon_{u'u}=\Upsilon_{uu'}^{-1}$, and therefore the matrices $\Upsilon_{uu'}$, $uu'\in \vec E^*$, form an $\su$-connection on $G^*$.
If this connection is \emph{flat}, meaning that the product of the matrices along any closed directed loop in $G^*$ (called the \emph{holonomy} of this loop) is the identity matrix, then one can 
consistently recover all $\xi_u$ from any fixed $\xi_{u_0}$. To prove Theorem~\ref{thm:main} it is therefore enough
to show that $\{ \Upsilon_{uu'} \}_{uu' \in \vec E^*}$ is a flat connection.

\subsection{Euler angles and rotations in 3D}
To this end, we will need a geometric interpretation of $\su$ matrices. 
Recall that the \emph{Euler angles} $\phi$ and $\psi$ associated with $U$ as in \eqref{eq:su2} are given by
\[
\psi + \phi=2 \arg a \qquad \textnormal{and} \qquad \psi -\phi = 2\arg b.
\]
A computation yields that for $\Upsilon_{uu'}$, the Euler angles are
\[
\psi_{uu'}=\beta_{uu'}-\frac \pi2 \qquad \textnormal{and} \qquad \phi_{uu'} =\frac \pi 2- \alpha_{uu'}-\beta_{uu'}=\frac {3\pi}2-\beta_{u'u},
\]
where we again used \eqref{eq:angeq}. Hence we can decompose into elemental $\su$ matrices:
\begin{align}
\Upsilon_{uu'}&=\begin{pmatrix}
e^{\frac i2(\psi_{uu'}+\phi_{uu'})}\cos(\frac{\vartheta_{uu'}}2) & e^{\frac i2(\psi_{uu'}-\phi_{uu'})}\sin(\frac{\vartheta_{uu'}}2)  \\
-e^{-\frac i2(\psi_{uu'}-\phi_{uu'})}\sin(\tfrac{\vartheta_{uu'}}2)  & e^{-\frac i2(\psi_{uu'}+\phi_{uu'})}\cos(\tfrac{\vartheta_{uu'}}2)
\end{pmatrix}\nonumber \\
&=
\begin{pmatrix} e^{i \frac{\psi_{uu'}}2}&  0 \\
0 & e^{-i \frac{\psi_{uu'}}2}
\end{pmatrix}
\begin{pmatrix}
\cos(\frac{\vartheta_{uu'}}2) & \sin(\frac{\vartheta_{uu'}}2)  \\
-\sin(\tfrac{\vartheta_{uu'}}2)  & \cos(\tfrac{\vartheta_{uu'}}2)
\end{pmatrix}
\begin{pmatrix} e^{i \frac{\phi_{uu'}}2}&  0 \\
0 & e^{-i \frac{\phi_{uu'}}2} 
\end{pmatrix} \nonumber \\
&=:\Upsilon_z(\psi_{uu'})\Upsilon_x(\vartheta_{uu'})\Upsilon_z(\phi_{uu'}). \label{eq:exdec}
\end{align}

The notation $\Upsilon_x(\gamma)$ and $\Upsilon_z(\gamma)$ refers to rotations around the $x$ and $z$ axis of $\mathbb R^3$ by an angle $\gamma$.
To justify it, we need to recall the connection between $\su$ and $\textnormal{SO}(3)$ -- the group of rotations of $\mathbb R^3$. We chose to use the representation of $\su$ as the unit \emph{quaternions}, i.e.
each $\Upsilon$ as in \eqref{eq:su2} is associated with the quaternion $q_{\Upsilon}=a+b\qi+c\qj+d\qk$. With the standard rules of quaternion multiplication 
($\qi^2=\qj^2=\qk^2=\qi\qj\qk=-1$, $\qi\qj=\qk$, $\qj\qk=\qi$ and $\qk\qi=\qj$)
the map $\Upsilon \mapsto q_{\Upsilon}$ is a group isomorphism 
between $\su$ and quaternions $q=a+b\qi+c\qj+d\qk$ of unit norm, i.e. such that $\|q\|^2=a^2+b^2+c^2+d^2=1$. 
Recall that $\bar q= a-b\qi-c\qj-d\qk$ and $q\bar q=\bar q q =\|q\|^2$. We call a quaternion $q$ \emph{pure} if $\Re q :=\tfrac12(q+\bar q)=0$. The set of pure quaternions can be identified with $\mathbb R^3$ by the map
 $b\qi+c\qj+d\qk \mapsto (d,c,b)$ (we chose this particular order for convenience so that $\qi$ corresponds to the $z$ coordinate and $\qk$ to the $x$ coordinate in $\mathbb R^3$). Each unit quaternion $q$ can be written as
 \begin{align*}
 q= \cos\Big(\frac \gamma 2\Big)+\sin\Big(\frac \gamma 2\Big) p
 \end{align*}
where $p$ is a pure unit quaternion. It is classical that if $p_0$ is another pure quaternion, then
 \begin{align*}
 p'_0=q p_0 q^{-1}
 \end{align*}
 is the pure quaternion resulting from the rotation of $p_0$ around the axis given by $p$ by an angle of $\gamma$ (note the doubling of the angle). Here the rotation is clockwise when looking in the direction of $p$.
 This map $q \mapsto R_p(\gamma)$, where $R_p(\gamma)\in \textnormal{SO}(3)$ is the corresponding rotation matrix, is a 2-to-1 group homomorphism from $\su$ to $\textnormal{SO}(3)$.
 The multiplicity of this map comes from the fact that $-q$ gives rise to the same rotation as $q$. This will be enough knowledge about $\su$ to finish the proof of Theorem~\ref{thm:main}.
 In particular, this justifies the notation from \eqref{eq:exdec}. 
 
 Before proceeding we still need to recall one classical fact, this time about rotations themselves.
 Namely, a sequence of rotations around the coordinate axes of a fixed reference frame (\emph{extrinsic rotations}) is equivalent to the reverse sequence of the same rotations but this time taken around the 
 corresponding axes of the moving (rotating) frame (\emph{intrinsic rotations}). This follows from the fact that the change of basis operation is realised by matrix conjugation (here the change of basis matrices are the rotations themselves).
 By \eqref{eq:exdec}, $\Upsilon_{uu'}$ corresponds to the rotation resulting from the composition $R_z(\psi_{uu'})R_x(\vartheta_{uu'})R_z(\phi_{uu'})$,
 where $x$ (resp.~$z$) stands for $e_1$ (resp.~$e_3$) in the standard frame $(e_1,e_2,e_3)$. Assuming the initial frame agrees with the standard frame, this corresponds to the composition
\begin{align} \label{eq:intseq}
R_{z''}(\phi_{uu'})R_{x'}(\vartheta_{uu'})R_z(\psi_{uu'}),
\end{align}
where $x'=R_z(\psi_{uu'})e_1$ and $z''=R_{x'}(\vartheta_{uu'})R_z(\psi_{uu'})e_3$, are the $x$ and $z$ axes of the rotating frame after one and two rotations respectively. 
This interpretation will be very convenient to show that the connection $\{\Upsilon_{uu'}\}_{uu'\in \vec G^*}$ is flat. Before doing this, we will need one more definition.

\subsection{Framed immersions and flatness of the connection}
\begin{figure}
        \includegraphics[scale=0.20]{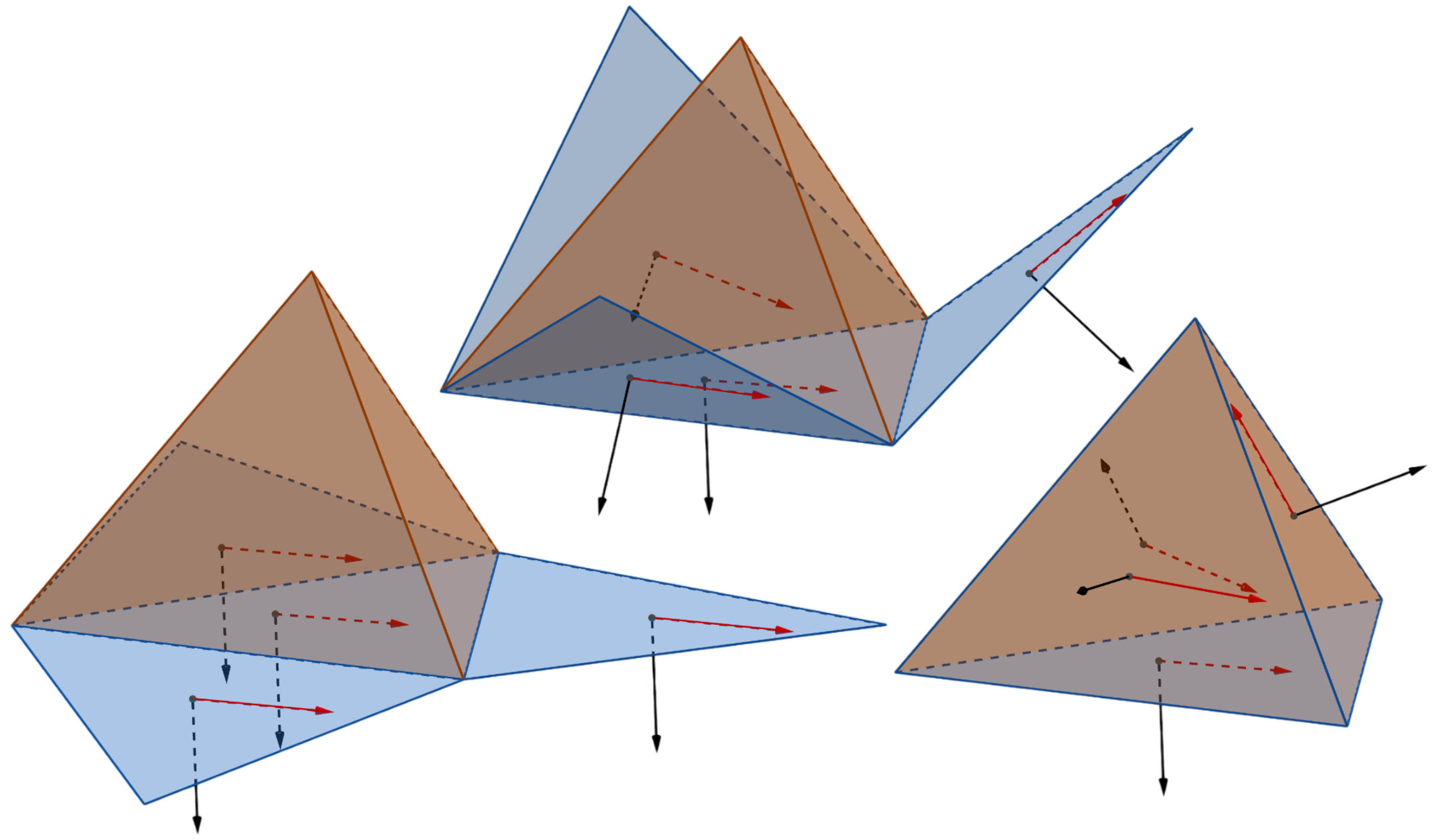} 
    \caption{From left to right: the folding of a net (a special case of a flat decomposition) of a tetrahedron together with a constant vector field (red vectors) on the faces into a framed immersion of the tetrahedron. 
    To match our conventions, the bottom plane on which the net lies should be seen upside down so that the black vectors point upwards.
    The red vectors in the plane are in our case taken to be~$i$, and they get mapped to the $\mathbf r_u$ vectors by the folding}
    \label{fig:framedim}
\end{figure}

Recall that for a face $u=\{v_1,v_2,v_3\}$, $\mathbf p_u$ is the plane on which $\I_{v_1},\I_{v_2},\I_{v_3}$ lie. 
A \emph{framed immersion} is an oriented immersion $\I$ together with a choice of a unit vector $\mathbf r_u \in \mathbb R^3$ parallel to $\mathbf p_u$ for each face $u\in U$ 
(see Fig~\ref{fig:framedim} for an illustration).

We now describe how to obtain a framed immersion associated with a planar decomposition. Before reading the formal definition the reader may look at Fig.~\ref{fig:framedim} which 
visualises this procedure readily in the case of the tetrahedron. To be more formal, consider a planar decomposition $\F$ of an immersion $\I$. For each $u=\{v_1,v_2,v_3\}\in U$, let 
$\iota_u$ be the isometry of $\mathbb R^3$ that maps $\mathbf T^\F_u$ to $\mathbf T^\I_u$, and such that $\iota_u(\F_u+(0,0,1))-\iota_u(\F_u)=\n_u$.
Here we still think of $\mathbb C\simeq \mathbb R^2$ as the $x-y$ plane of $\mathbb R^3$. Then we simply define $\mathbf r_u= \iota_u(i)$, where the complex unit $i$ is identified with $(0,1,0)\in \mathbb R^3$.

\begin{figure}
        \includegraphics[scale=0.7]{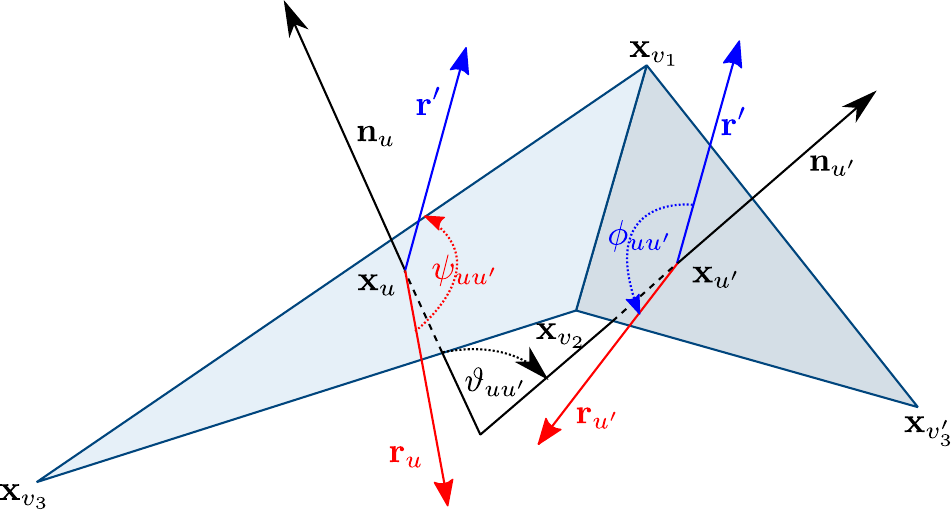}    
    \caption{Two adjacent faces in a framed immersion obtained from a planar decomposition.
    The composition of three intrinsic rotations $R_{x''}(\phi_{uu'})R_{z'}(\vartheta_{uu'})R_x(\psi_{uu'})$ rotates the frame $(\mathbf n_u,\mathbf r_u)$ to the frame $(\mathbf n_{u'},\mathbf r_{u'})$ by the sequence of transformations
    $(\mathbf n_u,\mathbf r_u)\to (\mathbf n_u,\mathbf r')\to (\mathbf n_{u'},\mathbf r')\to (\mathbf n_{u'},\mathbf r_{u'})$. The intermediate vector $\mathbf r'_u$ is equal to $(\I_{v_1}-\I_{v_2})/|\I_{v_1}-\I_{v_2}|$}
    \label{fig:intrrot}
\end{figure}

In this setup we now make the following claims (that are visualised in Fig.~\ref{fig:intrrot}). 
Assume without loss of generality that the initial frame is $(\mathbf r_u, \mathbf n_u\times \mathbf r_u, \mathbf n_u)=(e_1,e_2,e_3)$ (we will only care about $(\mathbf r_u,\mathbf n_u)$ as the third axis is determined)
and consider the sequence of intrinsic rotations from \eqref{eq:intseq}:
\begin{align*} 
R_{z''}(\phi_{uu'})R_{x'}(\vartheta_{uu'})R_z(\psi_{uu'}).
\end{align*}
We have that
\begin{itemize}
\item $R_z(\psi_{uu'})=R_{\mathbf n_u}(\psi_{uu'})$ maps $(\mathbf r_u,\mathbf n_u)$ to $(\mathbf r',\mathbf n_u)$, where 
\[
\mathbf r' =\frac{\I_{v_1}-\I_{v_2}}{|\I_{v_1}-\I_{v_2}|}.
\]
Indeed 
\begin{align*}
\psi_{uu'} &= \beta_{uu'}-\frac \pi2 =\arg\Big(\frac{\F_{uv_1}-\F_{uv_2}}{i}\Big)
\end{align*}
is the rotation angle from $i$ to $\F_{uv_1}-\F_{uv_2}$, which by construction is the same as the rotation angle from $\mathbf r_u$ to $\mathbf r' $.
\item $R_{x'}(\vartheta_{uu'})=R_{\mathbf r'}(\vartheta_{uu'})=R_{\I_{v_1}-\I_{v_2}}(\vartheta_{uu'})$ maps $(\mathbf r',\mathbf n_u)$ to $(\mathbf r',\mathbf n_{u'})$
by the definition of the angle $\vartheta_{uu'}$.
\item $R_{z''}(\phi_{uu'})=R_{\mathbf n_{u'}}(\phi_{uu'})$ maps $(\mathbf r'_u,\mathbf n_{u'})$ to $(\mathbf r_{u'}, \mathbf n_{u'})$. Indeed
\begin{align*}
\phi_{uu'}& =\pi +\frac {\pi}2-\beta_{u'u} =\pi +\arg\Big(\frac{i}{\F_{u'v_2}-\F_{u'v_1}}\Big) =\arg\Big(\frac{i}{\F_{u'v_1}-\F_{u'v_2}}\Big) 
\end{align*}
is the rotation angle from $\F_{u'v_1}-\F_{u'v_2}$ to $i$, which by construction is the same as the rotation angle from $\mathbf r'$ to $\mathbf r_{u'}$.
\end{itemize}

This means that $\Upsilon_{uu'}$ acts by rotating the frame at $u$ to the frame at $u'$. Hence if we follow a closed loop starting at $u$ and apply $\Upsilon$ iteratively, the frame at $u$ will be mapped to itself.
However, since the homomorphism from $\su$ to $\textnormal{SO}(3)$ is 2-to-1 it is still possible that the {holonomy} of the this loop is equal to minus the identity matrix (which also corresponds to the trivial rotation). 
 
We will now argue why this is not the case. It is enough to show that the holonomy of elementary loops defining the faces of $G^*$ is the identity (loops going around a single vertex $v$ of $G$).
To this end note that in the flat case, i.e.~when all the $\vartheta$ angles along the loop are zero, a planar decomposition can be chosen so that the paths between $uu'$ and $u'u$ have zero winding $\alpha_{uu'}$ (indeed they are trivial paths if $\F_{uu'}=\F_{u'u}$, or can be chosen to make two opposite right turns). In that case by \eqref{eq:angeq} we have $e^{\beta_{uu'}-\beta_{u'u}}=-1$, and hence $e^{\psi_{uu'}+\varphi_{u'u}}=1$, which in turn 
implies that $\Upsilon_{uu'}=\Id$ for all edges $uu'$ along the loop. In particular the product over the loop is the identity. One can now modify the picture by 
continuously changing the angles~$\vartheta$ and $\varphi$ around $v$. The holonomy along the loop is clearly a 
continuous function of all the angles (as seen in~\eqref{eq:exdec}), and hence it cannot discontinuously change from $\Id$ to $-\Id$. This ends the proof of Theorem~\ref{thm:main}.

\begin{remark}
As mentioned in the introduction, from the proof it follows that any choice of $\xi_{u_0}$ for some fixed $u_0\in U$ gives rise to an eigenvector of $\Lambda$ with eigenvalue one 
(by propagating $\xi_{u_0}$ to any other face by applications of $\Upsilon$). It also follows that any such eigenvector gives rise to such $\xi_{u_0}$, and hence the kernel of $\Id-\Lambda$ has 
two complex dimensions corresponding to the two coordinates of~$\xi_{u_0}$.
\end{remark}

\begin{remark}
The collection $\{\xi_{u}\}_{u\in U}$ can be seen as a field of two-component \emph{spinors} acted upon by the $\su$ connection $\{\Upsilon_{uu'}\}_{uu'\in \vec E^*}$. 
Each such field represents a null eigenvector of $\Id-\Lambda$.
\end{remark}

\subsection*{Acknowledgments}
I am grateful to the Institute for Pure and Applied Mathematics, UCLA for its hospitality during the program \emph{Geometry, Statistical Mechanics, and Integrability},
during which this research was commenced. I am also grateful to the Institute Mittag-Leffler for its hospitality during the conference \emph{Quantum Fields and Probability II} 
during which this research advanced considerably. 
This research was supported by FWF Standalone grant P 36298 \emph{Spins, Loops and Fields} 
and the SFB F 1002 grant \emph{Discrete Random Structures}.

\bibliographystyle{amsplain}
\bibliography{3d}
\end{document}